\documentclass[preprint,11pt,3p]{elsarticle}
\usepackage[T1]{fontenc}
\usepackage[english]{babel}
\usepackage{hyperref}
\usepackage{ifpdf}
\usepackage{subfigure}
\usepackage{amssymb}
\usepackage{amsfonts}
\usepackage{epsf}
\usepackage{rotating}
\usepackage{graphicx}
\usepackage{amsmath}
\usepackage{fancyhdr}
\usepackage{lineno}
\usepackage{dutchcal}
\usepackage{babel}
\usepackage{graphics}
\usepackage{pstricks}
\usepackage{color}
\usepackage{multirow}

\newcommand{\nn}{\nonumber}
\newcommand{\lsim}{\mathrel{\mathop{\kern 0pt \rlap
  {\raise.2ex\hbox{$<$}}}
  \lower.9ex\hbox{\kern-.190em $\sim$}}}
\newcommand{\gsim}{\mathrel{\mathop{\kern 0pt \rlap
  {\raise.2ex\hbox{$>$}}}
  \lower.9ex\hbox{\kern-.190em $\sim$}}}

\newcommand{\be}{\begin{equation}}
\newcommand{\ee}{\end{equation}}
\newcommand{\bea}{\begin{eqnarray}}
\newcommand{\eea}{\end{eqnarray}}

\begin{document}

\begin{frontmatter}

  \title{\boldmath {\bf Exploring Scalar and Vector Bileptons at the LHC\\
  in a 331 Model}}

\author[label1]{Gennaro Corcella}
\address[label1]{INFN, Laboratori Nazionali di Frascati,
  \\Via E.~Fermi 40, 00044, Frascati (RM), Italy}
\ead{gennaro.corcella@lnf.infn.it}

\author[label5]{Claudio Corian\`o}
\address[label5]{Dipartimento di Matematica e Fisica `Ennio De Giorgi',
  Universit\`a del Salento and INFN, Sezione di
  Lecce, \\ Via Arnesano, 73100 Lecce, Italy}
\ead{claudio.coriano@le.infn.it}

\author[label5]{Antonio Costantini}
\ead{antonio.costantini@le.infn.it}

\author[label5]{Paul H. Frampton}
\ead{paul.h.frampton@gmail.com}

\begin{abstract}
  We present an analysis on the production of two same-sign lepton pairs at the LHC, mediated by bileptons in the $SU(3)_c\times SU(3)_L\times U(1)_X$ theory, the so-called 331 model. Compared to other 331 scenarios, in this model the embedding of the hypercharge is obtained with the addition of 3 exotic quarks and doubly-charged vector gauge bosons with lepton numbers $\pm 2$ in the spectrum ($Y^{\pm\pm}$), which can mediate the production of four-lepton final states. Furthermore, a complete description of the model requires the introduction of a Higgs scalar sector, which is a sextet of $SU(3)_L$, necessary in order to correctly account for the lepton masses. As a result of this, new doubly-charged scalar states $H^{\pm\pm}$ are part of the spectrum as well and can in principle compete with the vector bileptons in giving rise to four-lepton final states. We investigate both channels and study several observables at the LHC, for both signal and Standard Model $ZZ$ background. With respect to previous work on
  vector- and scalar-bilepton production, we use the most updated exclusion
  limits at the LHC and implement the 331 model in a full Monte Carlo
  simulation code, capable of being interfaced with any analysis framework.
  Our result is that the
  331 signal can be discriminated from the background with a significance of 6-9 standard deviations, depending on the LHC luminosity, with the vector bileptons dominating over the scalar ones.

\end{abstract}

\end{frontmatter}

\section{Introduction}
In the experimental searches for new physics of the fundamental interactions
at the LHC, the resolution of several open issues - which remain 
unanswered within the Standard Model (SM) - typically
call for larger gauge structures and a wider particle content.
Such are the gauge-hierarchy problem in the Higgs sector or the origin of 
light-neutrino masses, just to mention two of them, both requiring
such extensions.
Grand Unified Theories (GUTs)  play certainly an important role in this:
unfortunately, Grand Unification needs an energy scale
${\cal O}(10^{12}$-$10^{15})$~GeV,
which is far higher than the electroweak one, currently probed at the LHC. 
Obviously, the process of identifying the main signatures of a certain
symmetry-breaking pattern from the GUT scale down to the TeV scale is
far from being trivial, due to the enlarged symmetries and to a parameter
space of considerable size. There are, however, interesting scenarios where
larger gauge symmetries can be discovered or ruled out already
at the TeV scale, which suggests some alternative paths of exploration.
Such is the case of a version of the $SU(3)_c\times SU(3)_L\times U(1)_X$ model,
also known as 331 model \cite{PHF,PP,Valle}, 
where the requirement that the gauge couplings are
real provides a significant upper bound on the physical region in which
its signal could be searched for. This property sets the vacuum
expectation values (vevs)
of the Higgs bosons, which trigger the breaking from the 331-symmetry scale
down to the electroweak one, around the TeV. 
The model that we consider allows for bileptons,
i.e. gauge bosons $(Y^{--}, Y^{++})$ of charge $Q=\pm 2$ and lepton number
$L=\pm 2$, and therefore we shall
refer to it as the {\em bilepton model}.
In the family of 331 models, bileptons in the spectrum are obtained only for
special embeddings of the $U(1)_X$ symmetry and of the charge $(Q)$ and hypercharge $(Y)$ generators in the local gauge structure. 
One additional feature of the model is that,
unlike the Standard Model or most chiral models formulated so
far,  extending the SM spectrum and symmetries, the number of (chiral)
fermion generations is underwritten by the cancellation of the gauge anomalies.
Gauge anomalies cancel among different fermion families, and select the
number of generations to be 3: from this perspective, the model appears to be
quite unique. Moreover, in the formulation of \cite{PHF}, which we shall
adopt hereafter, the third fermion
family is treated asymmetrically with respect to the first two families.

In a previous analysis \cite{cccf}
we have presented results for the production of
pairs of vector bileptons ($Y^{++}Y^{--}$) decaying into two same-sign
lepton pairs, in conjunction with two jets at the LHC, at
$\sqrt{s}=13$~TeV and relying on 
a full Monte Carlo implementation of the 331 model.
Our study has been based on the selection of a specific benchmark point in the
parameter space, where the $Y$ bileptons
have mass $m_Y\simeq 875$~GeV.
We have shown that, 
by setting appopriate cuts on the rapidities
and transverse momenta of the final-state leptons and jets,
it is possible to suppress the Standard Model background.
By including jets in the final state, one obtains 
larger signal/background ratio, but nonetheless one has to face 
the issue of jet reconstruction.

In this paper, we wish to extend the investigation in \cite{cccf}.
In fact, in \cite{cccf} the explicit mass matrices of the scalars
in a minimal version of the model, containing only 3 $SU(3)_L$ triplets, as well as the minimization conditions of the potential, were presented.
However, the model also allows for doubly-charged Higgs-like scalars
$(H^{++}, H^{--})$, which may give rise to multi-lepton final state,
in the same manner as the vectors $Y^{\pm\pm}$ debated in \cite{cccf}.
We shall therefore explore the production of same-sign lepton pairs
at the LHC, mediated by both scalar and vector bileptons;
unlike Ref.~\cite{cccf}, we shall veto final-state jets.

The production of doubly-charged vector bilepton pairs at the LHC in
jetless Drell--Yan
processes was already investigated in \cite{dion,barreto,alves,nepo}.
In particular, the authors of
\cite{nepo} implemented the bilepton model in a full Monte Carlo
simulation framework and obtained the exclusion limit
$m_Y^{\pm\pm}>850$~GeV, by using the ATLAS rates
at $\sqrt{s}=7$~TeV \cite{atlasold}
on expected and observed
high-mass same-sign lepton pairs and extending the results to
13 TeV and $\mathcal{L}=50~{\rm fb}^{-1}$.
However, the analysis in
\cite{nepo} assumes that the same-sign lepton-pair yields are
independent of the bilepton spin: in fact, the ATLAS analysis in
\cite{atlasold} was performed for scalar doubly-charged Higgs bosons,
while the predictions in \cite{nepo} concerned vector bileptons.
It is precisely the goal of the present exploration understanding
whether, referring to the 331 realization in \cite{PHF},
one can separate vector from scalar bileptons at the LHC,
in different luminosity regimes.
A careful investigation on the production of doubly-charged particles
at the LHC and the dependence of final-state distributions on the
spin was undertaken in \cite{fuks1}. The authors of \cite{fuks1}
considered an effective simplified model where the SM is extended by means of
a $SU(2)_L$ group and the scalar, fermion and vector doubly-charged
particles lie in the trivial, fundamental and adjoint representations
of $SU(2)_L$, respectively. By looking
at transverse-momentum and angular distributions
and varying the bilepton mass betwen 150 and 350 GeV, it was found that
it could be possible to distinguish at the LHC the particle spin.

Compared to our previous study, in the present paper
we shall explore a complete version of the
model which includes both the $SU(3)_L$ triplet 
Higgses and the newly added scalar sextet sector, which is necessary in order
to account for the masses of the leptons.
The inclusion of the scalar sextet opens up the decay channels
$H^{\pm\pm}\to l^\pm l^\pm$, which compete with the companion
$Y^{\pm\pm}\to l^\pm l^\pm$ process.
Many of the analytic expressions in the
description of the sextet contributions, such as the rotation 
matrices appearing in the extraction of the mass eigenstates of this sector,
cannot be given in closed analytical form, since they would be
too lengthy. As in \cite{cccf} we shall choose a benchmark point of
the model and present our results numerically: in particular,
as we are interested in comparing vector- and scalar-bilepton signals,
we shall set the doubly-charged $Y^{++}$ and $H^{++}$ masses to the same
value.

Vector- and scalar-bilepton production at hadron colliders
was also explored in \cite{ramirez}, where the authors presented the total
cross sections, expected number of events at the LHC, invariant-mass and
transverse-momentum spectra for a few values of the bilepton mass.
The decay properties of doubly-charged Higgs bosons in a minimal
331 model were also studied in \cite{tonasse}. It was found that,
since the coupling to leptons is proportional to the lepton mass, 
such scalar bileptons 
mostly decay into $\tau$-lepton pairs. Also, according to
\cite{tonasse}, the rate into $WW$ pairs is
suppressed, being proportional to the vacuum expectation value of
the Higgs giving mass to the neutrinos, as well as decays into leptons
of different flavours, since the Yukawa couplings are diagonal.
While the investigation in \cite{ramirez} was performed
at leading order, by using the \texttt{FORM} package \cite{form}
to calculate the ampitudes, we shall undertake a full hadron-level
investigation.
We will implement the 331 model, including
the sextet sector, 
into \texttt{SARAH 4.9.3} \cite{sarah}, while the amplitudes
for bilepton production at the LHC will be
computed by the \texttt{MadGraph} code \cite{madgraph};
the simulation of parton showers and hadronization 
will be performed by using \texttt{HERWIG} 6 \cite{herwig}.
Also, as will be thoroughly debated later on, in our
model doubly-charged scalar bileptons decay in all lepton-flavour
pairs with branching ratios 1/3, unlike Ref.~\cite{tonasse},
wherein the $\tau^+\tau^-$ mode had the largest rate.

From the experimental viewpoint, to our knowledge, there has been no
actual search for vector bileptons at the LHC,
whereas the latest investigations on possible
doubly-charged scalar Higgs boson production at the LHC were undertaken
in \cite{atlashh} and \cite{cmshh} by ATLAS and CMS, respectively.
In detail, the ATLAS analysis, performed at 13 TeV and with
$36~{\rm fb}^{-1}$ of data, considered the so-called left-right
symmetric model (LRSM, see, e.g. Refs.~\cite{pati1,pati2})
and its numerical implementation in \cite{spira},
where doubly-charged Higgs bosons can couple to either
left-handed or right-handed leptons. In this framework, assuming
that the $H^{\pm\pm}$ bosons only decay into lepton pairs,
exclusion limits were set in the range 770-870 GeV for
$H^{\pm\pm}_L$ and 660-760 GeV for $H^{\pm\pm}_R$.
As for CMS, a luminosity of $12.9~{\rm fb}^{-1}$ was taken into
account and limits between 800 and 820 were determined,
always under the assumption of a 100\% branching ratio into
same-sign lepton pairs.

Our paper is organized as follows. In Section 2, we shall
discuss the family embedding in the minimal 331 model, while Section 3 will
be more specific on its scalar content, giving details on
the triplet and sextet sectors, as well as on the lepton
masses and physical Higgs bosons.
Our phenomenological analysis will be presented in Section 5
and final comments and remarks will be given in Section 6.

\section{Family embedding in the minimal 331}
One of the main reasons
for the appearance of exotic particles in the spectrum of the 331 model
is the specific embedding of the 
hypercharge $Y$ in the $SU(3)_L\times U(1)_X$ gauge symmetry. The embeddings of $Y$ and of the charge operator 
$Q_{em}$ are obtained by defining them as linear combinations of the diagonal generators of $SU(3)_L$. We recall that in the 331 case this is defined by 
\begin{equation}
{Y}_{\bf 3} =\beta T_8 + X \mathbf{1} \qquad {Y}_{\bf \bar{3}} =-\beta T_8 + X \mathbf{1} 
\end{equation}
for the ${\bf 3}$ and the $\bar{\bf 3}$  representations
of $SU(3)_L$, respectively, with generators
$T_i=\lambda_i/2$ ($i=1,\ldots 8$), corresponding to the Gell-Mann matrices,
and $T_8=\textrm{diag}\left[ \frac{1}{2 \sqrt{3}}( 1,1,-2)\right]$.
The charge operator is given by 
\begin{equation} 
Q_{em, {\bf 3}}= Y_{\bf{3}} + T_3 \qquad Q_{em, \bar{\bf 3}}= Y_{\bar{\bf 3}} - T_3
\end{equation}
in the fundamental and anti-fundamental representations of 
of $SU(3)_L$, respectively, where we have $T_3$=${\rm diag}\left[ \frac{1}{2}
  (1,-1,0)\right]$.
We choose 
the $SU(2)_L\times U(1)_Y$ hypercharge assignments of the Standard Model
as $Y(Q_L)=1/6$, $Y(L)=-1/2$, $Y(u_R)=2/3$, $Y(d_R)=-1/3$ and $Y(e_R)=-1$.
Denoting by $q_X$ the particle charges
under $U(1)_X$, the breaking of the symmetry 
$SU(3)_L\times U(1)_X \to SU(2)_L\times U(1)_Y$, for the fundamental
representation {\bf 3} reads   
\begin{equation}
  ({\bf 3}, q_X) \to \left(2,\frac{\beta}{ 2 \sqrt{3}} +  q_X\right)
  +\left(1, -\frac{\beta}{\sqrt{3}} +  q_X\right),
\end{equation}
while for the representation ${\bf \bar{3}}$
\begin{equation}
  ({\bf\bar{3}}, q'_X) \to \left(2,-\frac{\beta}{2\sqrt{3}} +  q'_X\right)
  +\left(1, +\frac{\beta}{\sqrt{3}} +  q'_X\right).
\end{equation}
The $X$-charge
is fixed by the condition that the first two components of the
$Q_1$ and $Q_2$ triplets 
carry the same hypercharge 
as the quark doublets $Q_L=(u,d)_L$ of the Standard Model, yielding
\begin{equation}
\label{one}
q_X=\frac{1}{6} - \frac{\beta}{2\sqrt{3}}. 
\end{equation}
The $U(1)_{em}$ charge of the triplet will then be
$Q_{em}(Q_1)$=diag$(2/3,-1/3,1/6 - \sqrt{3}\beta/2)$.
Fermions with exotic charges will be automatically present in the case of
$\beta=\sqrt{3}$, which is the parameter choice that we will consider
herafter in our analysis. The first two families will then be assigned as 
\begin{equation}
Q_1=\left(
\begin{array}{c}
u_L\\
d_L\\
D_L
\end{array}
\right),\quad Q_2=\left(
\begin{array}{c}
c_L\\
s_L\\
S_L
\end{array}
\right),\quad Q_{1,2}\in({\bf 3},{\bf  3}, -1/3)
\end{equation}
under $SU(3)_c \times SU(3)_L \times U(1)_X$. The charge operator $Q_{em}$ on $Q_{1,2}$ will then give 
\begin{equation}
Q_{em}(Q_{1,2})=\textrm{diag} (2/3,-1/3,-4/3), 
\end{equation}
with two exotic quarks $D$ and $S$ of charge -4/3. The third family
is instead assigned as
\begin{equation}
Q_3=\left(
\begin{array}{c}
b_L\\
t_L\\
T_L
\end{array}
\right),\quad Q_3\in({\bf 3},{\bf \bar3}, 2/3), 
\end{equation}
where the hypercharge content of the third exotic quark ($T_L$) is derived
from the operator $(Y_{\bf \bar{3}})$ 
\begin{equation}
  Y_{\bf \bar{3}}(Q_3)= \textrm{diag}\left(-\frac{\beta}{2\sqrt{3}}+ q'_X,-
  \frac{\beta}{2\sqrt{3}}+ q'_X, \frac{\beta}{\sqrt{3}} + q'_X\right),
\end{equation}
with the condition $q'_X=1/6 +\beta/(2 \sqrt{3})$, giving 
\begin{equation}
Y_{\bf \bar{3}}(Q_3)= \textrm{diag}\left(\frac{1}{6},\frac{1}{6},\frac{5}{3}\right)
\end{equation}
and 
\begin{equation}
  Q_{em \,\bf \bar{3}}(Q_3)= \textrm{diag}\left(-\frac{1}{3},\frac{2}{3},\frac{5}{3}
  \right).
\end{equation}
With these assignments, the charge of $T_L$ is 
$Q_{em}(T_L)= 5/3$, allowing to distinguish between the third and the first
two generations of quarks.
Right-handed singlet quarks in the 331 model carry
the usual SM charges ($2/3$ and $-1/3$ for the $u$-type and
$d$-type quarks), 
\begin{align}
({d_R}, { s_R},{ b_R})&\in ({\bf  3}, 1,- 1/3)\\
( u_R,  c_R,  t_R) &\in({\bf  3}, 1, 2/3),\\
\end{align}
with the exception of the three right-handed exotics
\begin{align}
( D_R,  S_R) &\in ({\bf  3}, 1, -4/3)\\
 T_R &\in ({\bf 3}, 1, 5/3).
\end{align}
The lepton sector is assigned to the representation
$\bar{3}$ of the
same gauge group. Conversely from the quark sector, there is a democratic
arrangement of the three lepton generations into triplets of $SU(3)_L$, 
\begin{equation}\label{lee}
l=\left(
\begin{array}{c}
e_L\\
\nu_L\\
e_R^{\mathcal{c}}
\end{array}
\right),\quad l\in({\bf 1},{\bf \bar 3}, 0),\end{equation}
with ${e}_R^{\mathcal{c}}=i \sigma_2 e_R^*$.
In the following, we shall adopt for the leptons the notation
$l_a^i$, where the subscripts ($a$, $b$ or $c$)
refer to the lepton generation
(electrons, muons and taus), and the superscripts
($i,j,k=1,2,3$) are $SU(3)_L$ indices.
For example, the generation $a$ corresponds to electrons and the
three elements of the triplet (\ref{lee}) are labelled as:
\begin{equation}
l^1_a=e_{a\, L},\ \ 
  l^2_a=\nu_{a\, L},\ \ l^3_a=e_{a R}^{\mathcal{c}}. 
\end{equation}
For the hypercharge operator we have the decomposition under $SU(2)_L\times U(1)_Y$
\begin{equation} 
  Y_{\bf \bar{3}}(l)=\left(-\frac{\beta}{2 \sqrt{3}}+ q''_X,-
  \frac{\beta}{2 \sqrt{3}} + q''_X,\frac{\beta}{\sqrt{3}} + q''_X\right)
\end{equation} 
with $q''_X=1/6 +\beta/(2 \sqrt{3})$ and $Q_{em}(L)$=diag$(-1,0,1)$. Both left- and right-handed components of the SM
leptons are fitted into the same $SU(3)_L$ multiplet.

The scalars of the 331 model, responsible for the electroweak symmetry
breaking (EWSB), come in three triplets of $SU(3)_L$:
\begin{equation}
\rho=\left(
\begin{array}{c}
\rho^{++}\\
\rho^+\\
\rho^0
\end{array}
\right)\in(1,3,1),\quad\eta=\left(
\begin{array}{c}
\eta^+\\
\eta^0\\
\eta^-
\end{array}
\right)\in(1,3,0),\quad\chi=\left(
\begin{array}{c}
\chi^0\\
\chi^-\\
\chi^{--}
\end{array}
\right)\in(1,3,-1).
\end{equation}

The breaking $SU(3)_L\times U(1)_X\to U(1)_{em}$ is obtained in two steps.
The vacuum expectation value  of the neutral component of $\rho$ causes
the breaking from $SU(3)_L\times U(1)_X$ to $SU(2)_L\times U(1)_Y$;
the usual spontaneous symmetry breaking
mechanism from $SU(2)_L\times U(1)_Y$ to $U(1)_{em}$ is then
obtained through the vevs of the neutral components of $\eta$ and $\chi$. 

Before closing this section, we remind that in the models \cite{PHF,PP}
the coupling constants of $SU(3)_L$ and $U(1)_X$, namely $g_{3L}$ and
$g_X$, are related to the electroweak mixing angle $\theta_W$ in such
a way that $g_X(\mu)$ exhibits a Landau pole at a scale $\mu$ whenever
$\sin^2\theta_W(\mu)= 1/4$
\cite{landau1}.\footnote{The relation between the $SU(3)_L$ and $U(1)_X$
  couplings reads: $g_X^2/g_{3L}^2 =\sin^2 \theta_W / (1-4 \sin^2 \theta_W)$
\cite{landau1,landau2}.}
Therefore, 
the theory may lose its perturbativity even at the scale about 3.5 TeV
\cite{landau2}. Nevertheless, the typical energy scale of
bilepton-pair production in \cite{cccf} and in this paper
is somewhat smaller, i.e. $2m_{Y^{\pm\pm}}\simeq 1.75$~TeV.
Therefore, we shall assume that the Landau pole does not 
pose any threat to the perturbative analysis carried out in the
present work.

\section{The scalar sectors}
The model presented in the previous section exhibits
the interesting feature of having both scalar and vector doubly-charged
bosons, which is a peculiarity of the minimal version of the 331 model.
In fact it is possible to consider various versions of the $SU(3)_c\times SU(3)_L\times U(1)_X$ gauge symmetry, usually parametrized by $\beta$ \cite{other331}. We discuss the case of $\beta=\sqrt3$, corresponding to the minimal version presented here, leading to vector boson with electric charge equal to $\pm 2$.
Doubly-charged states are interesting by themselves
because they can have distinctive features in terms of allowed decay channels,
for example the production of same-sign lepton pairs.
In the context of the minimal 331 model there is an even more interesting possibility. In fact, one can test whether a same-sign lepton pair has been produced by either a scalar or a vector boson. As we are going to explain, this feature will also shed light on the presence of a higher representation of the $SU(3)_c\times SU(3)_L\times U(1)_X$ gauge group, namely the sextet.

\subsection{The triplet sector}
In the previous section we have seen that the EWSB mechanism is realised in the 331 model by giving a vev to the neutral component of the triplets $\rho$, $\eta$ and $\chi$. The Yukawa interactions for SM and exotic quarks
are obtained by means of these scalar fields and are given by: 
\begin{align}
\mathcal{L}_{q, triplet}^{{Yuk.}}&=\big(y_d^1\; Q_1  \eta^*  d_R + y_d^2\; Q_2  \eta^*  s_R + y_d^3\; Q_3  \chi\,  b_R^*\nn\\
&\quad + y_u^1\; Q_1  \chi^*  u_R^* + y_u^2\; Q_2  \chi^*  c_R^* + y_u^3\; Q_3  \eta\,  t_R^*\\
&\quad + y_E^1\; Q_1\,  \rho^*  D_R^* + y_E^2\; Q_2\,  \rho^*  S_R^* + y_E^3\; Q_3\,  \rho\,  T_R^*\big) + \rm{h.c.},\nn
\end{align}
where $y^i_{d}$, $y^i_u$ and $y^i_E$ are the Yukawa couplings for down-,
up-type and exotic quarks, respectively. The masses of the exotic quarks are related to the vev of the neutral component of $\rho=(0,0,v_\rho)$ via the invariants
\begin{eqnarray}
 Q_1\,  \rho^*  D_R^*, Q_1\,  \rho^*  S_R^*&\sim & (3,3,-1/3)\times (1,\bar{3},-1)\times (\bar{3},1,4/3) \nonumber \\
 Q_3\,  \rho  T_R^* &\sim& (3,\bar{3},2/3)\times (1,{3},1)\times (\bar{3},1,-5/3), 
\end{eqnarray}
responsible of the breaking $SU(3)_c\times SU(3)_L\times U(1)_X \to SU(3)_c\times SU(2)_L\times U(1)_Y$. It is clear that, being
$v_\rho\gg v_{\eta,\chi}$, the masses of the exotic quarks are
${\cal O}(\rm{TeV})$ whenever the relation $y_E^i\sim1$ is satisfied.

\subsection{The sextet sector}
The need for introducing a sextet sector can be summarised as follows. 
A typical Dirac mass term for the leptons in the SM
is associated with the operator $\bar{l}_LH e_R$, with $l_L=(v_{eL},e_L)$ being the $SU(2)_L $ doublet, with the representation content $(\bar{2},1/2)\times (2,1/2)\times(1,-1)$ (for $l, H$ and $e_R$, respectively) in $SU(2)_L\times U(1)_Y$. In the 331 the $L$ and $R$ components of the lepton $(e)$ are in the same multiplet and therefore the identification of an $SO(1,3)\times SU(3)_L$ singlet needs two leptons in the same representation. It can be obtained (at least in part) with the operator
\begin{eqnarray}
\mathcal{L}_{l,\, triplet}^{Yuk}&=& G^\eta_{a b}( l^i_{a \alpha}\epsilon^{\alpha \beta} l^j_{b \beta})\eta^{* k}\epsilon^{i j k} + \rm{h. c.}\nn\\
&=& G^\eta_{a b}\, l^i_{a}\cdot l^j_{b}\,\eta^{* k}\epsilon^{i j k} + \rm{h. c.} 
\end{eqnarray}
where the indices $a$ and $b$ run over the three generations of flavour, $\alpha$ and $\beta$ are Weyl indices contracted in order to generate  an $SO(1,3)$ invariant ($l^i_{a}\cdot l^j_{b}\equiv l^i_{a \alpha}\epsilon^{\alpha \beta} l^j_{b \beta}$) from two Weyl fermions, and $i,j,k=1,2,3$, are $SU(3)_L$ indeces.  The use of 
$\eta$ as a Higgs field is mandatory, since the components of the multiplet $l^j$ are $U(1)_X$ singlets. 
The representation content of the operator $l^i_a l^j_b$ according to $SU(3)_L$ is given by $3\times 3= 6 + \bar{3}$, with the 
$\bar{3}$ extracted by an anti-symmetrization over $i$ and $j$ via $\epsilon^{i j k}$. This allows to identify 
$l^i_a l^j_b \eta^{*k}\epsilon^{i j k}$ as an $SU(3)_L$ singlet. Considering that the two leptons are anticommuting Weyl spinors, and that the $\epsilon^{\alpha\beta}$ (Lorentz) and $\epsilon^{i j k}$ ($SU(3)_L$) contractions introduce two sign flips under the $a\leftrightarrow b$ exchange, the combination 
\begin{equation}
M_{a b}=( l^i_{a}\cdot l^j_{b })\eta^{* k}\epsilon^{i j k} 
\end{equation}
is therefore antisymmetric under the exchange of the two flavours, implying
that even $G_{a b}$ has to be antisymmetric.  However, an antisymmetric $G^\eta_{a b}$ is not sufficient to provide mass to all leptons.

In fact, the diagonalization of $G^\eta$ by means a unitary matrix 
$U$, namely $G^\eta=U \Lambda U^\dagger$, with $G^\eta$ antisymmetric in
flavour space, implies that its 3 eigenvalues are given by
$\Lambda=(0,\lambda_{22}, \lambda_{33})$, with $\lambda_{22}=-\lambda_{33}$,
i.e. one eigenvalue is null and the other two are equal in magnitude.
At the minimim of $\eta$, i.e. $\eta=(0,v_\eta,0)$, one has:
  \begin{equation} 
G^\eta_{a b}M^{a b}=-Tr (\Lambda\, U M U^\dagger)=
2 v_{\eta}\lambda_{22}\, U_{2 a}\, l^{1}_{a}\cdot l^{3}_{b}\,U_{2 b}^* + 2 v_{\eta}\lambda_{33}\, U_{3 a}\, l^{1}_{a}\cdot l^{3}_{b}\,U_{3 b}^*, 
 \end{equation}
with $ l^1_a=e_{a L}$ and  $l^3_b=e_{b R}^\mathcal{c}$. Introducing the linear combinations 
\begin{equation} 
  E_{2 L}\equiv U_{2 a}\, l^1_a=U_{2 a} \, '\, e_{a L} \qquad U_{2 b}^*\,
  l^3_b=U_{2 b}^*\, e_{b R}^\mathcal{c} = i\sigma_2(U_{2 b} \,e_{b R})^*\equiv
  E_{2 R}^\mathcal{c},
\end{equation}
then the antisymmetric contribution in flavour space becomes 
\begin{equation}
\mathcal{L}_{l, \, triplet}^{Yuk}= 2 v_{\eta}\lambda_{22} \left( E_{2 L}E_{2 R}^\mathcal{c}  - E_{3 L}E_{3 R}^\mathcal{c}\right),
\end{equation}
which is clearly insufficient to generate the lepton masses of three
non-degenerate
lepton families. We shall solve this problem by introducing a second 
invariant operator, with the inclusion of a sextet $\sigma$:
\begin{equation}
\sigma=\left(
\renewcommand*{\arraystretch}{1.5}
\begin{array}{ccc}
\sigma_1^{++}&\sigma_1^+/\sqrt2&\sigma^0/\sqrt2\\
\sigma_1^+/\sqrt2&\sigma_1^0&\sigma_2^-/\sqrt2\\
\sigma^0/\sqrt2&\sigma_2^-/\sqrt2&\sigma_2^{--}
\end{array}
\right)\in(1,6,0),
\end{equation}
leading to the Yukawa term
\begin{equation}\label{lag}
\mathcal{L}_{l, sextet}^{{Yuk.}}= G^\sigma_{a b} l^i_a\cdot l^j_b \sigma^*_{i,j},
\end{equation}
which allows to build a singlet out of the representation
$6$ of $SU(3)_L$, contained in $l^i_a\cdot l^j_b$, by combining it with
the flavour-symmetric
$\sigma^*$, i.e. $\bar{6}$.
Notice that $G^\sigma_{a b}$ is symmetric in flavour space. 

It is interesting to note that if one did not consider the sextet it would not
be possible for a doubly-charged scalar to decay into same-sign leptons.
In fact, if we leave aside the sextet contribution,
the Yukawa for the leptons is related to the scalar triplet $\eta$
which does not possess any doubly-charged state.
This means that revealing a possible decay
$H^{\pm\pm}\to l^\pm l^\pm$ would be a
distinctive signature of the presence of the sextet representation
in the context of the 331 model.

\subsection{Lepton Mass Matrices}
The lepton mass matrices are of course related to the Yukawa interactions by the Lagrangian 
\begin{equation}
\mathcal{L}_{l}^{{Yuk.}}=\mathcal{L}_{l, sextet}^{{Yuk.}} + \mathcal{L}_{l, triplet}^{{Yuk.}} + \rm{h. c.}
\end{equation}
and are combinations of triplet and sextet contributions.
The structure of the mass matrix that emerges
from the vevs of the neutral components of  $\eta$ and $\sigma$ is thus
given by:
\begin{equation} 
\mathcal{L}_{l}^{{Yuk.}}=\left(\sqrt{2} \sigma_0 G_{a,b}^\sigma  +2 v_\eta G^\eta_{a b}\right) (e_{a L}\cdot e_{b R}^\mathcal{c}) 
+\sigma_1^0 G^\sigma_{a b} \left(\nu_L^T i \sigma_2 \nu_L\right)  +\rm{h. c.},
\end{equation}
which generates a Dirac mass matrix for the charged leptons $M_{ab}^l$
and a Majorana mass matrix for neutrinos $M_{a b}^{\nu_l}$:
\begin{equation}\label{mlgen}
M^l_{a b} =\sqrt{2} \langle\sigma_0\rangle\, G_{a,b}^\sigma  +2 v_\eta\, G^\eta_{a b} \qquad , \qquad M_{a b}^{\nu_l}=\langle \sigma^0_1\rangle \, G^\sigma_{a b}. 
\end{equation}
In the expression above $\langle\sigma^0\rangle$ and $\langle\sigma_1^0\rangle$ are the vacuum expectation values of the neutral components of $\sigma$. For a vanishing $G^\sigma$, as we have already discussed, we will not be able to generate the lepton masses consistently, nor any mass for the neutrinos, i.e.   
\begin{equation}\label{mltip}
M^l_{a b}=2 v_\eta\, G^\eta_{a b} \qquad , \qquad M^{\nu_l}=0.
\end{equation} 
On the contrary, in the limit $G^\eta\to0$, Eq.~(\ref{mlgen}) becomes 
\begin{equation}\label{mlsext}
  M^l_{a b}=\sqrt{2} \langle\sigma_0\rangle\, G_{ab}^\sigma \qquad,
  \qquad M^{\nu_l}_{a,b}=\frac{\langle\sigma_1^0\rangle}{\sqrt2}G^\sigma_{a b},
\end{equation}
which has some interesting consequences.
Since the Yukawa couplings are the same for both leptons and neutrinos,
we have to require $\langle\sigma_1^0\rangle\ll\langle\sigma^0\rangle$, in order to obtain small neutrino masses. For the goal of our analysis, we will
assume that the vev of $\sigma_1^0$ vanishes, i.e.
$\langle\sigma_1^0\rangle\equiv0$. Clearly,
if the matrix $G^\sigma$ is diagonal in flavour space, from Eq.
(\ref{mlsext}) we will
immediately conclude that the Yukawa coupling $G^\sigma$ has
to be chosen to be proportional to the masses of the SM leptons.
An interesting consequence of this is that the decay $H^{\pm\pm}\to l^\pm l^\pm$, which is also proportional to $G^\sigma$, and therefore to the lepton masses,
will be enhanced for the heavier leptons, in particular for the $\tau$,
as thoroughly discussed in \cite{tonasse}.
This is an almost unique situation which is not encountered in other models with doubly-charged scalars decaying into same-sign leptons \cite{spira}.

However, for the sake of generality,
in the following analysis we will consider the most generic scenario where
both contributions $G^\sigma$ and $G^\eta$ are present. In this case the
branching ratios of the doubly-charged Higgs decaying into same-sign leptons
do not have to be proportional to the masses of the lepton anymore.
In particular, after accounting for both $G^\sigma$ and $G^\eta$,
configurations wherein even scalar bileptons have the same rates
into the three lepton species are allowed, as occurs for vector
bileptons.
In the following, we shall hence concentrate our investigation
on scenarios yielding 
\begin{equation}\label{br13}
  {\rm BR}(Y^{\pm\pm}\to l^\pm l^\pm)\simeq {\rm BR}(H^{\pm\pm}\to l^\pm l^\pm)
\simeq 1/3
\end{equation}
for $l=e,\mu,\tau$. The condition in Eq.~(\ref{br13})
is in fact particularly suitable to compare vector- and scalar-bilepton
rates at the LHC and, for the time being, 
should be seen
as part of our model. The assumption of having equal branchings of 1/3
in the decay of the scalar to all three lepton families allows
to extend the universality of spin-1 bileptons
also to the scalar sector, allowing to treat the two states
(scalar and vector) on a similar footing.
This option is clearly possible since we have 9 total parameters
in the mass matrix which are constrained by 6 conditions.
Three of them are necessary in order to reproduce the
lepton masses and the remaining three come from the requirement of having
equal values of the branching ratios of the scalars into the three lepton
families.
The explicit expressions of the solutions of such conditions
are very involved and we have hence opted for a numerical scanning of the mass
matrices satisfying such requirements.
This in general requires that the ratio of the matrix elements of $G_{\eta}$ over those of $G_\sigma$ to be proportional to $v_\sigma/v_\eta \sim 10^{-2}$.
Of course, if possible experimental
data deviated significantly from ${\rm BR}( l^\pm l^\pm)=1/3$,
then they would clearly favour a scalar bilepton, because
lepton-flavour universality is mandatory for vector bileptons.

\subsection{Physical Higgs bosons}
The inclusion of the sextet representation in the potential enriches the phenomenology of the model and enlarges the number of physical states in the spectrum. In fact we now have, after electroweak symmetry breaking (EWSB) $SU(3)_L\times U(1)_X \to SU(2)_L\times U(1)_Y\to   U(1)_{\rm{em}}$,
five scalar Higgses, three pseudoscalar Higgses, four charged Higgses and three doubly-charged Higgses. The (lepton-number conserving) potential of the model is given by \cite{TullyJoshi} 
\begin{align}\label{pot}
V&= m_1\, \rho^\dagger\rho+m_2\,\eta^\dagger\eta+m_3\,\chi^\dagger\chi +\lambda_1 (\rho^\dagger\rho)^2+\lambda_2(\eta^\dagger\eta)^2+\lambda_3(\chi^\dagger\chi)^2+\lambda_{12}\rho^\dagger\rho\,\eta^\dagger\eta\\
&\quad+\lambda_{13}\rho^\dagger\rho\,\chi^\dagger\chi+\lambda_{23}\eta^\dagger\eta\,\chi^\dagger\chi+\zeta_{12}\rho^\dagger\eta\,\eta^\dagger\rho+\zeta_{13}\rho^\dagger\chi\,\chi^\dagger\rho+\zeta_{23}\eta^\dagger\chi\,\chi^\dagger\eta\nn\\
&\quad + m_4\,Tr(\sigma^\dagger \sigma) + \lambda_{4} (Tr(\sigma^\dagger\sigma))^2 + \lambda_{14}\rho^\dagger\rho\,Tr(\sigma^\dagger\sigma) + \lambda_{24}\eta^\dagger\eta\,Tr(\sigma^\dagger\sigma) + \lambda_{34}\chi^\dagger\chi\,Tr(\sigma^\dagger\sigma) \nn\\
&\quad + \lambda_{44}Tr(\sigma^\dagger\sigma\,\sigma^\dagger\sigma)+ \zeta_{14} \rho^\dagger \sigma\,\sigma^\dagger \rho + \zeta_{24} \eta^\dagger\sigma\,\sigma^\dagger\eta + \zeta_{34} \chi^\dagger\sigma\,\sigma^\dagger\chi\nn\\
&\quad + (\sqrt2 f_{\rho\eta\chi} \epsilon^{ijk}\rho_i\, \eta_j\, \chi_k + \sqrt2 f_{\rho\sigma\chi} \rho^T\, \sigma^\dagger\, \chi \nn\\
&\quad+ \xi_{14}\epsilon^{ijk}\, \rho^{*l} \sigma_{li} \rho_j \eta_k + \xi_{24}\epsilon^{ijk}\epsilon^{lmn}\,\eta_i\eta_l\sigma_{jm}\sigma_{kn} + \xi_{34}\epsilon^{ijk}\,\chi^{*l}\sigma_{li}\chi_j\eta_k) + \rm{h.c.}\nn
\end{align}

The EWSB mechanism will cause a mixing among the Higgs fields;
from Eq. (\ref{pot}) it is possible to obtain the explicit expressions of the mass matrices of the scalar, pseudoscalar, charged and doubly-charged Higgses, by using
standard procedures. In the broken Higgs phase, the minimization conditions
\be\label{mincond}
\frac{\partial V}{\partial v_\phi}=0, \quad \langle \phi^0\rangle=v_\phi, \quad \phi=\rho, \eta, \chi, \sigma
\ee
will define the tree-level vacuum. We remind that we are considering massless neutrinos choosing the neutral field $\sigma_1^0$ to be inert. The explicit expressions of the minimization conditions are then given by
\begin{align}\label{minpot1}
m_1 v_\rho + \lambda_1 v_\rho^3 + \frac{1}{2}\lambda_{12}v_\rho v_\eta^2-f_{\rho\eta\chi} v_\eta v_\chi+\frac{1}{2}\lambda_{13}v_\rho v_\chi^2 - \frac{1}{\sqrt2}\xi_{14}v_\rho v_\eta v_\sigma + f_{\rho\sigma\chi}v_\chi v_\sigma&\\
+ \frac{1}{2}\lambda_{14}v_\rho v_\sigma^2 + \frac{1}{4}\zeta_{14}v_\rho v_\sigma^2&=0\nn\\
m_2 v_\eta + \frac{1}{2}\lambda_{12}v_\rho^2 v_\eta +\lambda_2 v_\eta^3 - f_{\rho\eta\chi} v_\rho v_\chi +\frac{1}{2}\lambda_{23} v_\eta v_\chi^2 - \frac{1}{2\sqrt2}\xi_{14}v_\rho^2 v_\sigma+ \frac{1}{2\sqrt2}v_\chi^2 v_\sigma&\\
+\frac{1}{2}\lambda_{24} v_\eta v_\sigma^2-\xi_{24} v_\eta v_\sigma^2&=0\nn\\
m_3 v_\chi + \lambda_3 v_\chi^3 + \frac{1}{2} \lambda_{13} v_\rho^2 v_\chi - f_{\rho\eta\chi} v_\rho v_\eta +\frac{1}{2}\lambda_{23}v_\eta^2 v_\chi +\frac{1}{\sqrt2}\xi_{34}v_\eta v_\chi v_\sigma + f_{\rho\sigma\chi} v_\rho v_\sigma&\\
+\frac{1}{2}\lambda_{34} v_\chi v_\sigma^2 + \frac{1}{4}\zeta_{34} v_\chi v_\sigma^2&=0\nn\\
\label{minpot2}
m_4 v_\sigma + \frac{1}{2}\lambda_{14}v_\rho^2 v_\sigma + \lambda_{44} v_\sigma^3 + \frac{1}{2}\lambda_4 v_\sigma^3 + f_{\rho\sigma\chi} v_\rho v_\chi - \frac{1}{2\sqrt2} \xi_{14} v_\rho^2 v_\eta + \frac{1}{2\sqrt2} \xi_{34} v_\eta v_\chi^2&\\
+\frac{1}{2}\lambda_{14}v_\rho^2 v_\sigma + \frac{1}{4}\zeta_{14} v_\rho^2 v_\sigma + \frac{1}{2}\lambda_{24} v_\eta^2 v_\sigma - \xi_{24} v_\eta^2 v_\sigma + \frac{1}{2} \lambda_{34} v_\chi^2 v_\sigma + \frac{1}{4} \zeta_{34} v_\chi^2 v_\sigma&=0
\nn
\end{align}
These conditions are inserted into the the tree-level mass matrices of the CP-even and CP-odd Higgs sectors, derived from $M_{ij}=\left.{\partial^2 V}/{\partial \phi_i\partial \phi_j}\right|_{vev},$, where $V$ is the potential in
Eq.~(\ref{pot}): the explicit expressions of the mass matrices
are too cumbersome to be presented here, although their calculation is rather straightforward. After a numerical diagonalization, we derive both the mass eigenstates and the Goldstone bosons.

In this case we have 5 scalar Higgs bosons, one of them will be the SM
Higgs with mass about 125 GeV, along with 4 neutral pseudoscalar Higgs bosons, out of which 2 are the Goldstones of the $Z$ and the $Z^\prime$ massive vector bosons. In addition there are 6 charged Higgses, 2 of which are the charged Goldstones and 3 are doubly-charged Higgses, one of which is a Goldstone boson. The Goldstones are exactly 8, as the massive vector bosons below the electroweak scale.

Hereafter we shall give the schematic expression of the physical Higgs states,
after EWSB, in terms of the gauge eigenstates, whose expressions contain only the vev of the various fields.
 In the following equations, ${\rm R}_{ij}^K\equiv{\rm R}_{ij}^K(m_1,m_2,m_3,\lambda_1,\lambda_2,\ldots)$ refers to the rotation matrix of each Higgs sector that depends on all the parameters of the potential in Eq.~(\ref{pot}). Starting from the scalar (CP-even) Higgs bosons we have
\bea
H_i = {\rm R}_{i1}^S {\rm Re}\, \rho^0 + {\rm R}_{i2}^S {\rm Re}\, \eta^0 +
{\rm R}_{i3}^S {\rm Re}\, \chi^0 +{\rm R}_{i4}^S {\rm Re}\, \sigma^0 +
{\rm R}_{i5}^S {\rm Re}\, \sigma_1^0,
\eea 
expressed in terms of the rotation matrix of the scalar components
${\rm R}^S$. There are similar expressions for the pseudoscalars 
\bea
A_i = {\rm R}_{i1}^P {\rm Im}\, \rho^0 + {\rm R}_{i2}^P {\rm Im}\,
\eta^0 +{\rm R}_{i3}^P{\rm Im}\, \chi^0 +{\rm R}_{i4}^P {\rm Im}\, \sigma^0 +{\rm R}_{i5}^P {\rm Im}\, \sigma_1^0.
\eea 
in terms of the rotation matrix of the pseudoscalar components ${\rm R}^P$.
Here, however, we have two Goldstone bosons responsible for the generation of the masses of the neutral gauge bosons $Z$ and $Z^\prime$ given by
\begin{align}
A_0^1 &= \frac{1}{N_1}\left(v_\rho {\rm Im}\,\rho^0 -v_\eta {\rm Im}\,\eta^0 + v_\sigma {\rm Im}\,\sigma^0\right),\,\qquad N_1=\sqrt{v_\rho^2+v_\eta^2+v_\sigma^2}\ ;\\
A_0^2 &= \frac{1}{N_2}\left(-v_\rho {\rm Im}\,\rho^0+ v_\chi {\rm Im}\,\chi^0\right),\,\qquad N_2=\sqrt{v_\rho^2+v_\chi^2}.
\end{align}

For the charged Higgs bosons the interaction eigenstates are
\bea
H_i^+ = {\rm R}^{C}_{i1}\rho^{+} + {\rm R}^{C}_{i2}(\eta^{-})^* + {\rm R}^{C}_{i3}\eta^{+} + {\rm R}^{C}_{i4}(\chi^{-})^* + {\rm R}_{i5}^C \sigma_1^+ + {\rm R}_{i6}^C (\sigma_2^-)^*
\eea 
with ${\rm R}^C$ being a rotation matrix of the charged sector. 
Even in this case we have two Goldstones because in the 331 model there are the $W^\pm$ and the $Y^\pm$ gauge bosons. The explicit expressions of the Goldstones are
\begin{align}
H_{W}^+ &= \frac{1}{N_W} \left(-v_\eta\eta^+ + v_\chi (\chi^-)^* + v_\sigma (\sigma_2^-)^*\right),\,\qquad N_W=\sqrt{v_\eta^2 + v_\chi^2 + v_\sigma^2}; \\
H_{Y}^+ &= \frac{1}{N_Y} \left(v_\rho \rho^+ - v_\eta (\eta^-)^* + v_\sigma \sigma_1^+\right),\,\qquad N_Y=\sqrt{v_\rho^2 + v_\eta^2 + v_\sigma^2}.
\end{align}
In particular, we are interested in the doubly-charged Higgses, where the number of physical states, after EWSB, is three, whereas 
we would have had only
one physical doubly-charged Higgs if we had not included the
sextet. The physical doubly-charged Higgs states are expressed in
terms of the gauge eigenstates and the elements of the
rotation matrix ${\rm R}^C$ as
\bea
H_i^{++} ={\rm R}^{2C}_{i1}\rho^{++} + {\rm R}^{2C}_{i2}(\chi^{--})^* + {\rm R}^{2C}_{i3}\sigma_1^{++} + {\rm R}^{2C}_{i4}(\sigma_2^{--})^*.
\eea
In particular, the structure of the corresponding Goldstone boson is
\bea
H_0^{++} =\frac{1}{N}\left(-v_\rho\rho^{++} + v_\chi(\chi^{--})^* - \sqrt2v_\sigma\sigma_1^{++} + \sqrt2v_\sigma(\sigma_2^{--})^*\right)
\eea
where $N=\sqrt{v_\rho^2+v_\chi^2+4v_\sigma^2}$ is a normalization factor.

\subsection{Vertices for $H^{\pm\pm}$ and $Y^{\pm\pm}$}

\begin{figure}[t]
\centering
\mbox{\subfigure[]{
\includegraphics[width=0.225\textwidth]{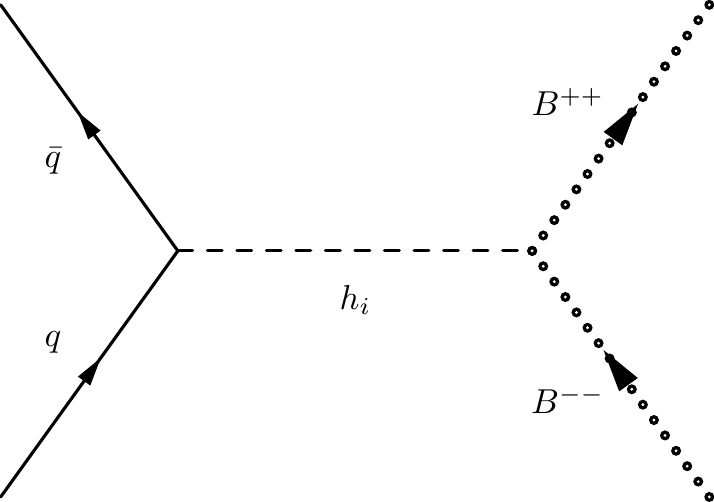}}\hspace{.8cm}
  \subfigure[]{
  \includegraphics[width=0.225\textwidth]{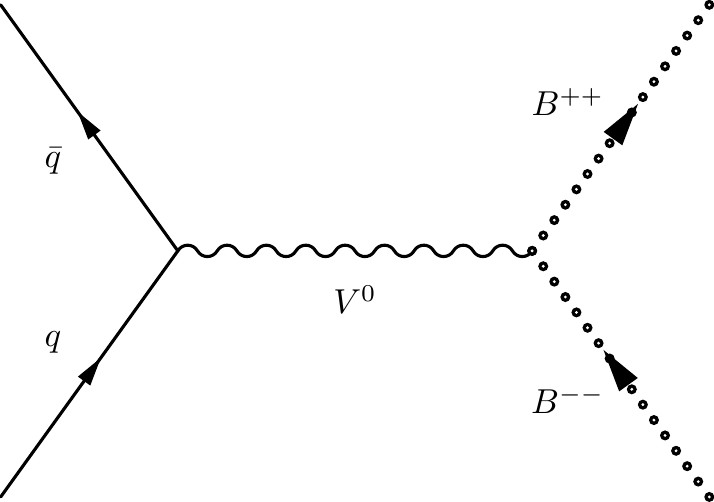}}\hspace{.8cm}
\subfigure[]{\includegraphics[width=0.225\textwidth]{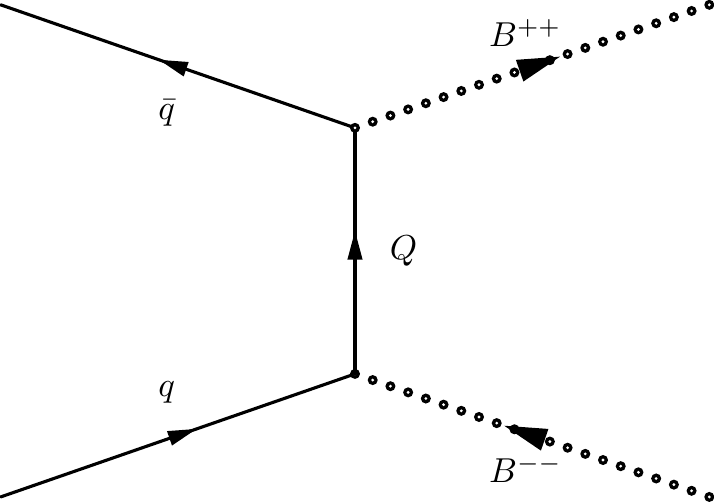}}}
\caption{Typical contributions to events with two doubly-charged bosons in the
  final state and no extra jets.
  (a) and (b): contributions due
  to the mediation of a scalar (a) and a vector (b) boson.
  (c): $t$-channel exchange of exotic quarks $Q$.}
 \label{jetless}
\end{figure}

In Fig.~\ref{jetless} we present the typical contributions to the
partonic cross section of the process $p p\to B^{++}B^{--}$, where $B^{\pm\pm}$
denotes either a spin-0 or a spin-1 bilepton;
each $B^{\pm\pm}$ decays into a same-sign lepton pair.
From Fig.~\ref{jetless}, we learn that bilepton pairs can be produced 
in Drell--Yan processes mediated by either a vector boson
($V^0=\gamma,Z,Z'$) or a scalar neutral Higgs ($h_1\cdots h_5$);
moreover, their production can be mediated by the exchange of
an exotic quark $Q$ in the $t$-channel as well. In principle, we may even have
$BB$ production via an effective vertex in gluon-gluon fusion, but
this contribution turned out to be negligible
with respect to the subprocesses with initial-state quarks.

In the following, we wish to discuss the differences between the
couplings of scalar Higgses and vector bosons to
scalar and vector bileptons, as the production
rates at the LHC crucially depend on such couplings.
Considering first the case of vector $Y^{\pm\pm}$,
the Lorentz structure of the
$V^0(p^1)Y^{++}(p^2)Y^{--}(p^3)$ vertex is given in terms of the momenta by
\be
V(p_\mu^1,p_\nu^2,p_\rho^3)=g_{\mu\nu}(p^2_\rho-p^1_\rho) + g_{\nu\rho}(p^3_\mu-p^2_\mu) + g_{\mu\rho}(p^1_\nu-p^3_\nu).
\ee  
Characterizing the vector boson $V$ as photon, $Z$ or $Z'$, we obtain:
\begin{align}\label{vzeroyy}
\gamma_\alpha\,Y_\mu^{++}\,Y_\nu^{--} &= -2 i g_2 \sin\theta_W\; V(p_\alpha^\gamma, p_\mu^{Y^{++}}, p_\nu^{Y^{--}})\nn\\
Z_\alpha\,Y_\mu^{++}\,Y_\nu^{--} &= \frac{i}{2}g_2 (1-2\cos 2\theta_W)\sin\theta_W\; V(p_\alpha^Z, p_\mu^{Y^{++}}, p_\nu^{Y^{--}})\\
Z_\alpha^\prime\,Y_\mu^{++}\,Y_\nu^{--} &= -\frac{i}{2}g_2 \sqrt{12-9\sec^2\theta_W}\; V(p_\alpha^{Z^\prime}, p_\mu^{Y^{++}}, p_\nu^{Y^{--}}),\nn
\end{align}
where $\theta_W$ is the Weinberg angle.

In the case of the doubly-charged Higgs boson,
the situation is slightly different: in fact, the interaction $V^0\, H^{++}\, H^{--}$ is generated
after that the Higgses take a vev.
The Lorentz structure of the coupling will be of course proportional to
the difference of the momenta of the Higgs fields. Defining
$S\left(p_\mu^1,p_\mu^2\right)= p^1_\mu - p^2_\mu,$ 
we have
\begin{align}\label{vzerohh}
  \gamma_\alpha\,H_i^{++}\,H_j^{--}&= -i \sin\theta_W \Big[\left(g_2+g_1
    \sqrt{\cot^2\theta_W -3}\right)\left({\rm R}^{2C}_{i1}{\rm R}^{2C}_{j1}+{\rm R}^{2C}_{i2}{\rm R}^{2C}_{j2}\right)\nn\\
&\qquad\qquad\qquad+2g_2\left({\rm R}^{2C}_{i3}{\rm R}^{2C}_{j3}+{\rm R}^{2C}_{i4}{\rm R}^{2C}_{j4}\right)\Big] S\left(p^{H_i^{++}}_\alpha, p^{H_j^{--}}_\alpha\right)\nn\\
& =-2 i e \delta_{i j}S\left(p^{H_i^{++}}_\alpha, p^{H_j^{--}}_\alpha\right)\\
Z_\alpha \,H_i^{++}\,H_j^{--}&=\frac{i}{2}\sec\theta_W\Big\{\cos 2\theta_W\left(g_2+g_1\sqrt{\cot^2\theta_W-3}\right)-g_1\sqrt{\cot^2\theta_W-3}){\rm R}^{2C}_{i1}{\rm R}^{2C}_{j1}\nn\\
&\qquad\qquad-2\big[\left(g_2+g_1\sqrt{\cot^2\theta_W -3}\right)\sin^2\theta_W {\rm R}^{2C}_{i2}{\rm R}^{2C}_{j2} -g_2\cos2\theta_W{\rm R}^{2C}_{i3}{\rm R}^{2C}_{j3}
  \nn \\
&\qquad\qquad+2g_2\sin^2\theta_W{\rm R}^{2C}_{i4}{\rm R}^{2C}_{j4}\big]\Big\}S\left(p^{H_i^{++}}_\alpha, p^{H_j^{--}}_\alpha\right)\\
Z^\prime_\alpha \,H_i^{++}\,H_j^{--}&=\frac{i}{2}\frac{\sec^2\theta_W}{\sqrt{12-9\sec^2\theta_W}}\Big\{\left[3g_1\sqrt{\cot^2\theta_W -3}(\cos2\theta_W-1)+g_2(2\cos2\theta_W-1)\right]{\rm R}^{2C}_{i1}{\rm R}^{2C}_{j1}\nn\\
&\qquad\qquad+\left[3g_1\sqrt{\cot^2\theta_W -3}(\cos2\theta_W-1)+2g_2(2\cos2\theta_W-1)\right]{\rm R}^{2C}_{i2}{\rm R}^{2C}_{j2}\nn\\
&\qquad\qquad +2g_2(2\cos2\theta_W-1)\left({\rm R}^{2C}_{i3}{\rm R}^{2C}_{j3}+2{\rm R}^{2C}_{i4}{\rm R}^{2C}_{j4}\right)\Big\}S\left(p^{H_i^{++}}_\alpha, p^{H_j^{--}}_\alpha\right).\nn\\
\end{align}

The interactions shown in Eq.~(\ref{vzeroyy}) and Eq.~(\ref{vzerohh}) are clearly very different, both in their Lorentz structures and in their dependence on the parameters of the model; therefore, different decay rates
$V^0,h_i\to B^{++}B^{--}$ are to be expected, according to whether
$B$ is a scalar or a vector. 
It can be noticed
that the expressions of the coupling in $\gamma\, Y^{++} Y^{--}$, i.e.
$2g_2\sin\theta_W\equiv 2e$, is apparently very different from the $\gamma\, H^{++}H^{--}$ one,  but one can show that, after simplifications,
they turn out to be the same, as expected. 
The relevant vertices for vector bileptons are
\bea
\ell \;\ell \; Y^{++}=\left\{
\begin{array}{cl}
-\frac{i}{\sqrt2}g_2 \gamma^\mu& P_L\\
\frac{i}{\sqrt2}g_2 \gamma^\mu& P_R
\end{array}
\right.\label{lly}
\eea
\bea
\bar d \;T \; Y^{--}=\left\{
\begin{array}{cl}
-\frac{i}{\sqrt2}g_2 \gamma^\mu &P_L\\
0& P_R
\end{array}
\right.\label{dty}
\eea
\bea
\bar D \;u \; Y^{--}=\left\{
\begin{array}{cl}
\frac{i}{\sqrt2}g_2 \gamma^\mu &P_L\\
0& P_R
\end{array}
\right.
\eea
\bea
h_i\;Y^{++}Y^{--}=\frac{i}{2}g_2^2\left(v_\rho {\rm R}^S_{i1}+v_\chi{\rm R}^S_{i3}\right)
\eea
\bea
\gamma\;Y^{++}Y^{--}=-2i\,g_2\,\sin\theta_W
\eea
\bea
Z\;Y^{++}Y^{--}=\frac{i}{2}\,g_2\,(1-2\cos2\theta_W)\sec\theta_W
\eea
\bea
Z'\;Y^{++}Y^{--}=-\frac{i}{2}\,g_2\,\sqrt{12-9\sec^2\theta_W}.
\eea
In the equations above, $P_{L,R}$ are the usual left- and right-handed
projectors $P_{L,R}=(1\mp\gamma_5)/2$.

\section{Phenomenological analysis at the LHC}

In this section we wish to present a phenomenological analysis, aiming
at exploring possible scalar- or vector-bilepton signals at the LHC.

As in our previous work,
we choose a specific benchmark point, obtained after
scanning the parameter space
by employing the \texttt{SARAH 4.9.3} program and its UFO
\cite{ufo} interface.
In doing so, we make use of the analytical expression of
the mass matrices and of the minimization conditions of the
potential in Eqs.~(\ref{minpot1})--(\ref{minpot2}).
The mass eigenvalues are computed numerically,
after varying all the quartic couplings in Eq.~(\ref{pot}) between
-1 and 1 and the vacuum expectation value $v_\rho$, responsible
of the first breaking of the 331 model, between 2 and 4 TeV. 
Possible benchmark points are then chosen in such a way that
all particle masses are positive, the SM-like
Higgs boson has mass at about 125 GeV and one is 
consistent with the current LHC exclusion limits
on new physics scenarios.
Moreover, as discussed in \cite{cccf}, we require the couplings
of the lightest neutral Higgs boson to the Standard Model fermions and 
bosons to be the same as in the Standard Model, within 10\% accuracy.
For the sake of comparison, we also choose
the doubly-charged vectors and scalars to have roughly
the same mass and just above the present ATLAS and CMS exclusion limits
\cite{atlashh,cmshh} on doubly-charged Higgs bosons.. 
Furthermore, we are obviously
interested in enhancing possible 331-model signals.
Limiting ourselves to the Higgs and exotic sectors,  
the particle masses in our reference point are quoted in Table~\ref{bp}.
\begin{table}
\begin{center}
\renewcommand{\arraystretch}{1.4}
\begin{tabular}{|c|c|c|}
\hline\hline
\multicolumn{3}{|c|}{Benchmark Point}\\
\hline
\hline
$m_{h_1}=126.3$ GeV & $m_{h_2}=1804.4$ GeV& $m_{h_3}=2474.0$ GeV\\
\hline
$m_{h_4}=6499.8$ GeV & $m_{h_5}=6528.1$ GeV& \\
\hline
$m_{a_1}=1804.5$ GeV& $m_{a_2}=6496.0$ GeV& $m_{a_3}=6528.1$ GeV\\
\hline
$m_{h^\pm_1}=1804.5$ GeV& $m_{h^\pm_2}=1873.4$ GeV & $m_{h^\pm_3}=6498.1$ GeV \\
\hline
$m_{h^{\pm\pm}_1}=878.3$ GeV& $m_{h^{\pm\pm}_2}=6464.3$ GeV & $m_{h^{\pm\pm}_3}=6527.7$ GeV \\
\hline
$m_{Y^{\pm\pm}}=878.3$ GeV& $m_{Y^\pm}=881.8$ GeV & $m_{Z'}=3247.6$ GeV\\
\hline
$m_D=1650.0$ GeV & $m_S=1660.0$ GeV& $m_T=1700.0$ GeV\\
\hline
\hline
\end{tabular}
\caption{Benchmark point for our collider study, consistent with the $\sim 125$ GeV Higgs mass and the present exclusion limits on BSM physics.}\label{bp}
\end{center}
\end{table}\par
From Table~\ref{bp}, we learn that the 331 model, as expected,
after including the sextet sector, yields 5 neutral scalar
($h_1, \dots h_5$); 3 pseudoscalar ($a_1$, $a_2$ and $a_3$) and 3
singly-charged ($h^\pm_1$, $h^\pm_2$, $h^\pm_3$) Higgs bosons:
the lightest $h_1$ is SM-like, whereas
the others have mass between 1.8 and 6.5 TeV. In particular,
$h_2$ is roughly degenerate with $a_1$ and $h^\pm_1$, while $h_4$, $h_5$, $a_2$,
$a_3$, $h^\pm_2$ and $h^\pm_3$ have all mass about 6.5 TeV.
As for doubly-charged particles, both $Y^{\pm\pm}$ and $h_1^{\pm\pm}$ have
mass around 878 GeV, just above the current exclusion limit for doubly-charged
scalars, while the other scalars $h_2^{\pm\pm}$ and $h_3^{\pm\pm}$
are in the 6.5 TeV range and the singly-charged vector $Y^\pm$ is roughly as
heavy as the doubly-charged one.
In our scenario, doubly-charged vectors and scalars decay only into
lepton pairs, with branching ratio 1/3 for each lepton family
($ee$, $\mu\mu$ or $\tau\tau$).
The exotic quarks $D$, $S$ and $T$
in our reference point have instead mass between 1.65 and 1.70 TeV.
In principle, such exotic quarks can be produced in pairs at the LHC,
with cross sections in the range of 0.5-0.7~fb at 13 TeV and
0.8-1.1~fb at 14 TeV, and may deserve
a complete phenomenological analysis, especially in the high-luminosity
LHC phase. 
Nevertheless, in the present 
paper we prefer to concentrate ourselves on the bilepton
phenomenology and
defer a thorough investigation on the production and decays
of exotic quarks
of charge 4/3 and 5/3 to future work \cite{ccgpp}.

The $Z'$ boson deserves further comments. In \cite{dion} 
the relation
\begin{equation}\label{zpy}
  \frac{m_{Y^{++}}}{m_{Z'}}\simeq \frac{\sqrt{3-12\sin^2\theta_W}}{2\cos\theta_W}
\simeq 0.27 \end{equation}
was determined between $Z'$ and vector-bilepton masses, and in fact
in Table~\ref{bp} Eq.~(\ref{zpy}) is verified to a pretty
good accuracy.
Moreover, in our benchmark scenario the $Z'$ width is
almost 700 GeV and, as found out in \cite{Dumm} when exploring
$Z'$ bosons in 331 models, 
our $Z'$ is leptophobic.
Therefore, the searches for $Z'$ bosons carried out so far
by ATLAS \cite{atlaszp} and CMS \cite{cmszp},
which have set
exclusion limits around 4 TeV on their mass, cannot be directly applied to our
scenario, since such searches were mostly perfomed for
narrow resonances decaying into dilepton final states \footnote{See, e.g.,
  Ref.~\cite{araz} on how the $Z'$ exclusion limits are modified in
  leptophobic models.}.
In our reference point, the $Z'$ decays dominantly 
into $q\bar q$ pairs, amounting to almost 70\% of the total width,  and
has a significant branching ratio into $Y^{++}Y^{--}$ pairs,
about 14\%; its decay rate into doubly-charged scalars $h^{++}_1h^{--}_1$ 
is instead rather small, roughly 1\%.
Such a difference can be easily explained in terms of the particle
spins: the $Z'$ has spin 1 and therefore, in the decay into
$h^{++}_1h^{--}_1$, only the amplitude where the $Z'$ has zero helicity with
respect to the $h^{++}_1h^{--}_1$ axis contributes.
On the contrary, in a possible decay into vector states
$Z'\to Y^{++}Y^{--}$, amplitudes with helicity 0 and $\pm 1$
with respect to the $Y^{++}Y^{--}$ direction play a role.

In the following, we shall present results for the production of
two same-sign lepton pairs at the LHC, mediated by either
vector or scalar bileptons in the 331 model:
\be
pp\to Y^{++}Y^{--}(H^{++}H^{--})\to (l^+l^+)(l^-l^-),
\label{signal}
\ee
where $l=e,\mu$ and, for simplicity, we have denoted by $H^{\pm\pm}$
the lightest doubly-charged Higgs boson $h_1^{\pm \pm}$.
The amplitude of process (\ref{signal}) is generated by 
the \texttt{MadGraph} code \cite{madgraph}, matched with \texttt{HERWIG} 6
for shower and hadronization \cite{herwig}.
We have
set $\sqrt{s}=13$~TeV and chosen the
NNPDFLO1 parton distributions \cite{nnpdf}, which are the default sets in
\texttt{MadGraph}.

As in Ref.~\cite{cccf}, and along the lines of \cite{atlashh,cmshh},
we set the following acceptance cuts on the 
lepton transverse momentum ($p_T$),
rapidity ($\eta$) and invariant opening angle ($\Delta R$):
\be\label{cuts}
p_{T,l}>20~{\rm GeV}, \ 
|\eta_l|<2.5,\  \Delta R_{ll}>0.1.
\ee
We point out that, 
since our signal originates from the decay of particles
with mass almost 1 TeV, the final-state electrons and muons
will be pretty boosted, and therefore
the actual values of the cuts in (\ref{cuts}) 
are not really essential, especially the transverse-momentum cut.
\footnote{Our cuts are in fact a conservative choice of the
  so-called overlap-removal algorithm implemented by ATLAS to discriminate
  lepton and jet tracks \cite{overlap}.}
At 13 TeV LHC, 
after such cuts are applied, the LO cross sections, computed by 
\texttt{MadGraph}, read
\be\sigma(pp\to YY\to 4l)\simeq
4.3~{\rm fb}\ ;\ \sigma(pp\to HH\to 4l)\simeq 0.3~{\rm fb} .\ee
Once again, the difference in the cross sections can be explained
in terms of the spin of the intermediate bileptons.
In the centre-of mass frame, in fact, for scalar production,
only the matrix element
where the vector ($\gamma$, $Z$ and $Z'$) has helicity zero
with respect to the $H^{++}H^{--}$ direction contributes; for
decays into $Y^{++}Y^{--}$ final states also the $\pm 1$ helicity
amplitudes are to be taken into account.
For processes mediated by scalars ($h_i\to H^{++}H^{--}/Y^{++}Y^{--}$),
the vector final states has still more helicity options since
$Y^{++}$ and $Y^{--}$ can rearrange their helicities in a few different
ways to achieve angular-momentum conservation, i.e. a total vanishing
helicity in the centre-of-mass frame.
We therefore confirm the findings of Ref.~\cite{ramirez}, where a higher
cross section for vector-bilepton production with respect to the scalars
was obtained at 7 and 14 TeV.

As for the background, final states with
four charged leptons may occur through intermediate
$Z$-boson pairs:
\be
pp\to ZZ\to (l^+l^-)(l^+l^-).
\label{zz}
\ee
After setting the same cuts as in (\ref{cuts}),
the LO cross section of the process (\ref{zz}) is given by
\be
\sigma(pp\to ZZ\to 4l)\simeq 6.1~{\rm fb}.
\ee
In principle, within the backgrounds, one should also consider
SM Higgs-pair production ($hh$),
with $h\to l^+l^-$. However, because of the tiny
coupling of the Higgs boson to electrons and muons, such a background turns
out to be negligible.

Assuming a luminosity of 300 fb$^{-1}$, the number of same-sign
electron/muon pairs in processes mediated by $YY$, $HH$ and $ZZ$ are
$N(YY)\simeq 1302$, $N(HH)\simeq 120$, $N(ZZ)\simeq 1836$.
Defining the significance $s$ to discriminate a signal $S$
from a background $B$ as
\be
s=\frac{S}{\sqrt{B+\sigma_B^2}},
\ee
$\sigma_B$ being the systematic error on B, which
we estimate as $\sigma_B\simeq 0.1 B$, we find that
the $YY$ signal can be separate from the $ZZ$ background with a
significance $s\simeq 6.9$, while  $HH$ production is overwhelmed
by both Standard Model background
($s=0.6$) and possible vector-bilepton pairs ($s=0.9$).

At 14 TeV, the cross sections read:
$\sigma(YY)\simeq 6.0$~fb, $\sigma(HH)\simeq 0.4$~fb and
$\sigma(ZZ)\simeq 6.6$~fb, leading to
$N(YY)\simeq 17880$, $N(HH)\simeq 1260$ and
$N(ZZ)\simeq 19740$ events with 3000 fb$^{-1}$ of data.
Therefore, in the high-luminosity phase of the LHC,
one will be able to discriminate vector-like bileptons
from the background with a significance of about 9 standard deviations,
while one is still unable to distinguish doubly-charged
Higgses from $YY$ ($s=0.70$) or $ZZ$ ($s=0.64$) pairs.

\begin{figure}[ht]
\centering
\mbox{\subfigure[]{
\includegraphics[width=0.450\textwidth]{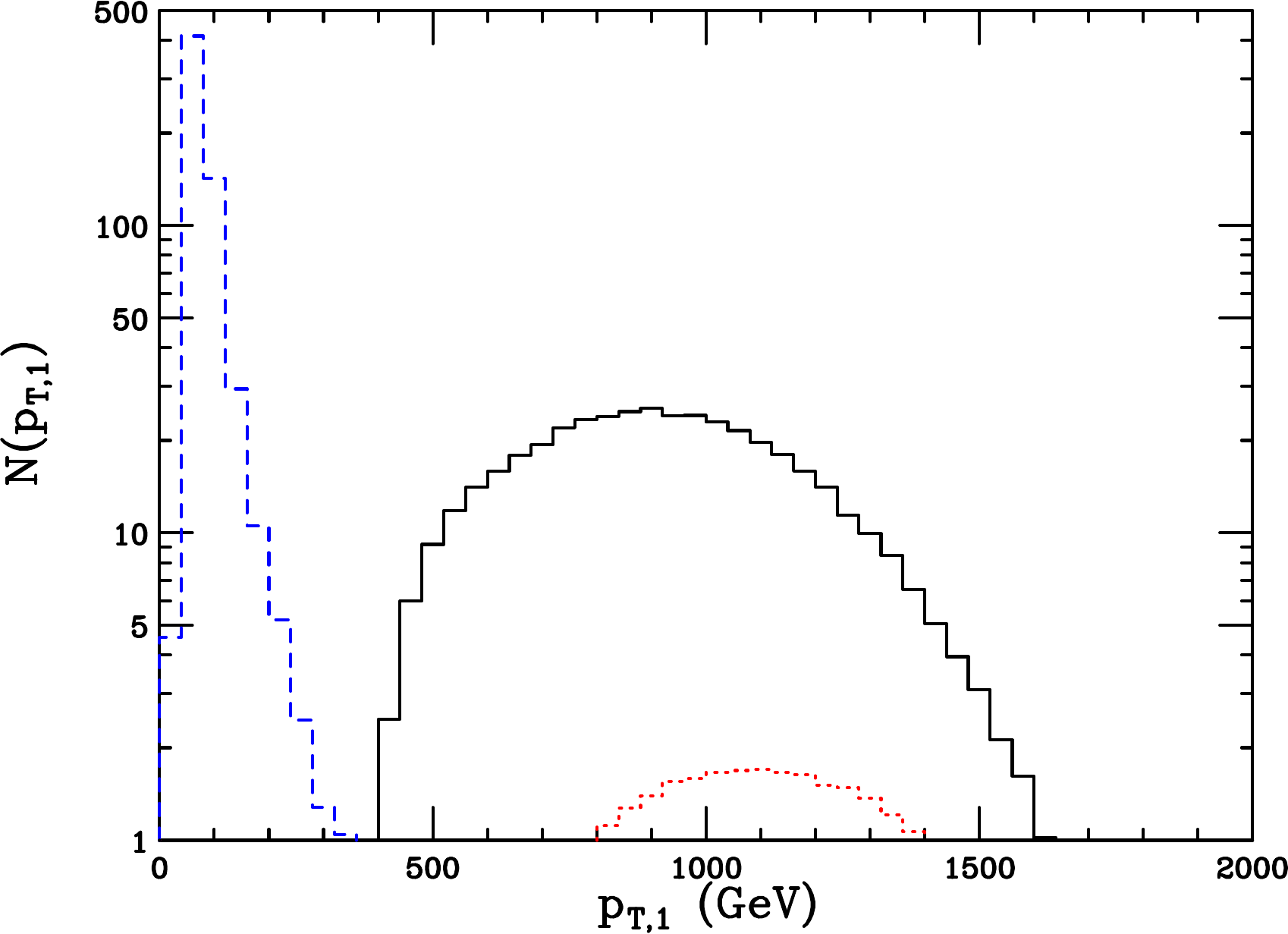}}\hspace{.1cm}
\subfigure[]{\includegraphics[width=0.450\textwidth]{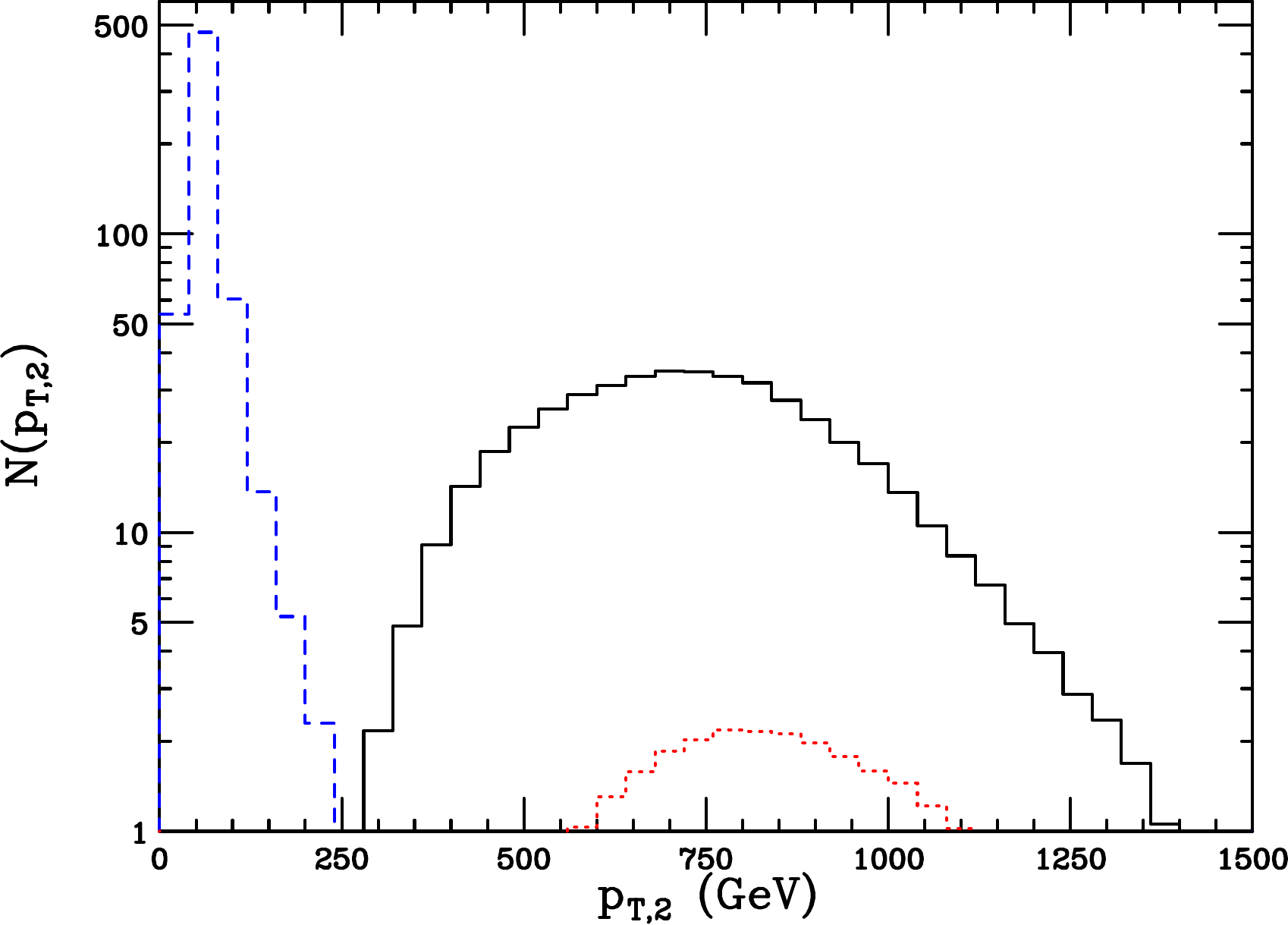}}}
\mbox{\subfigure[]{\includegraphics[width=0.450\textwidth]{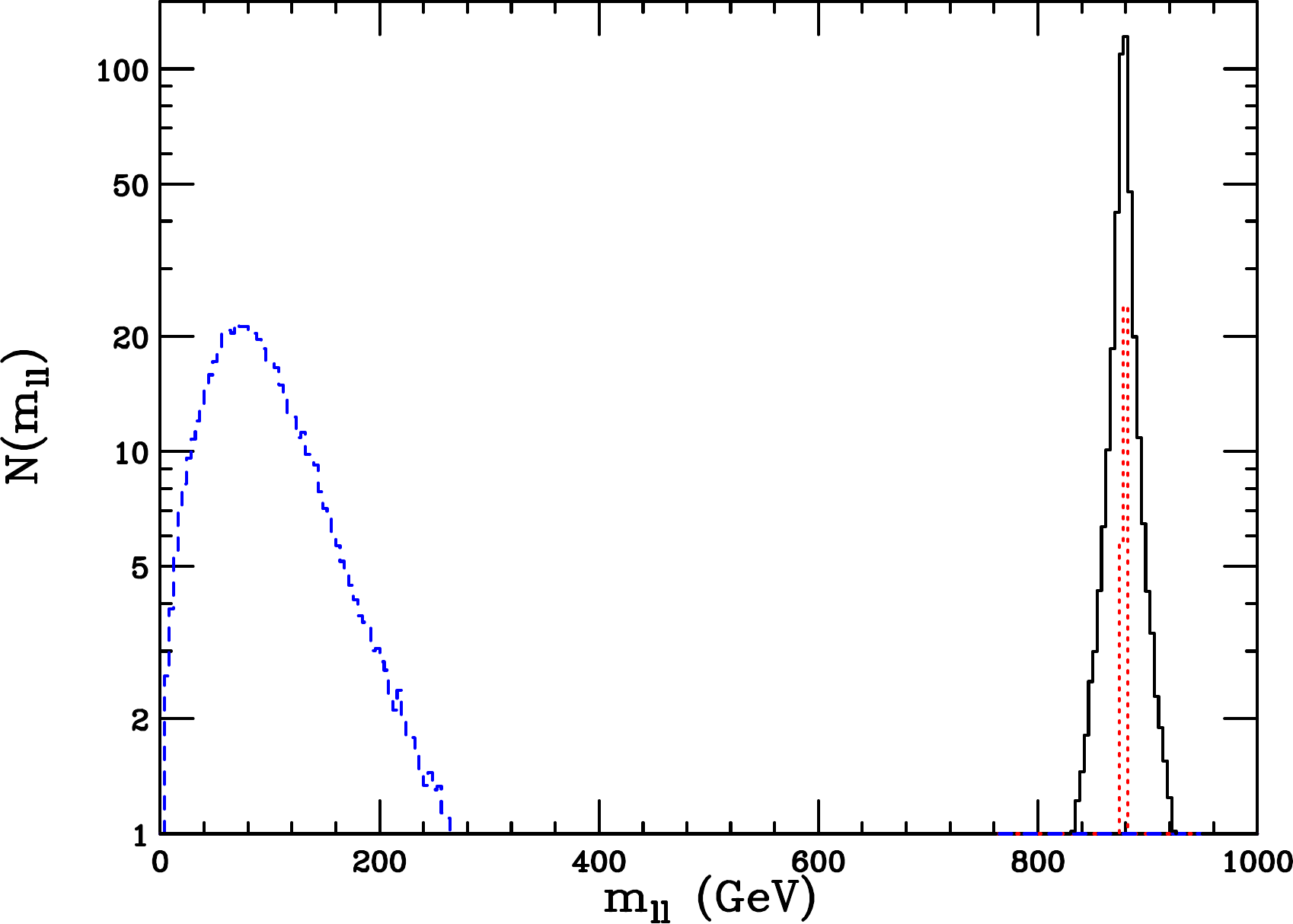}}\hspace{.1cm}
\subfigure[]{
\includegraphics[width=0.450\textwidth]{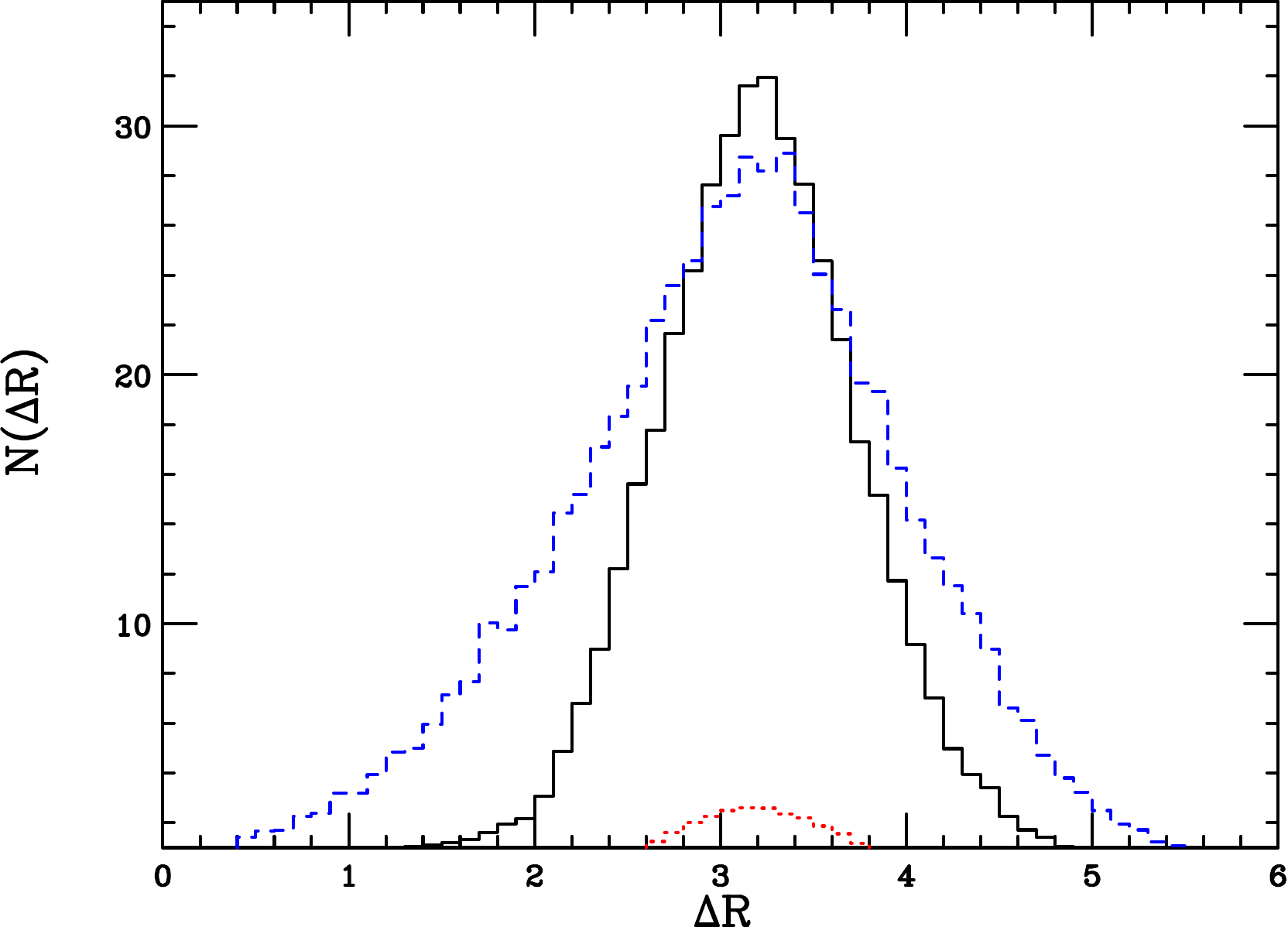}}}
\mbox{\subfigure[]{\includegraphics[width=0.450\textwidth]{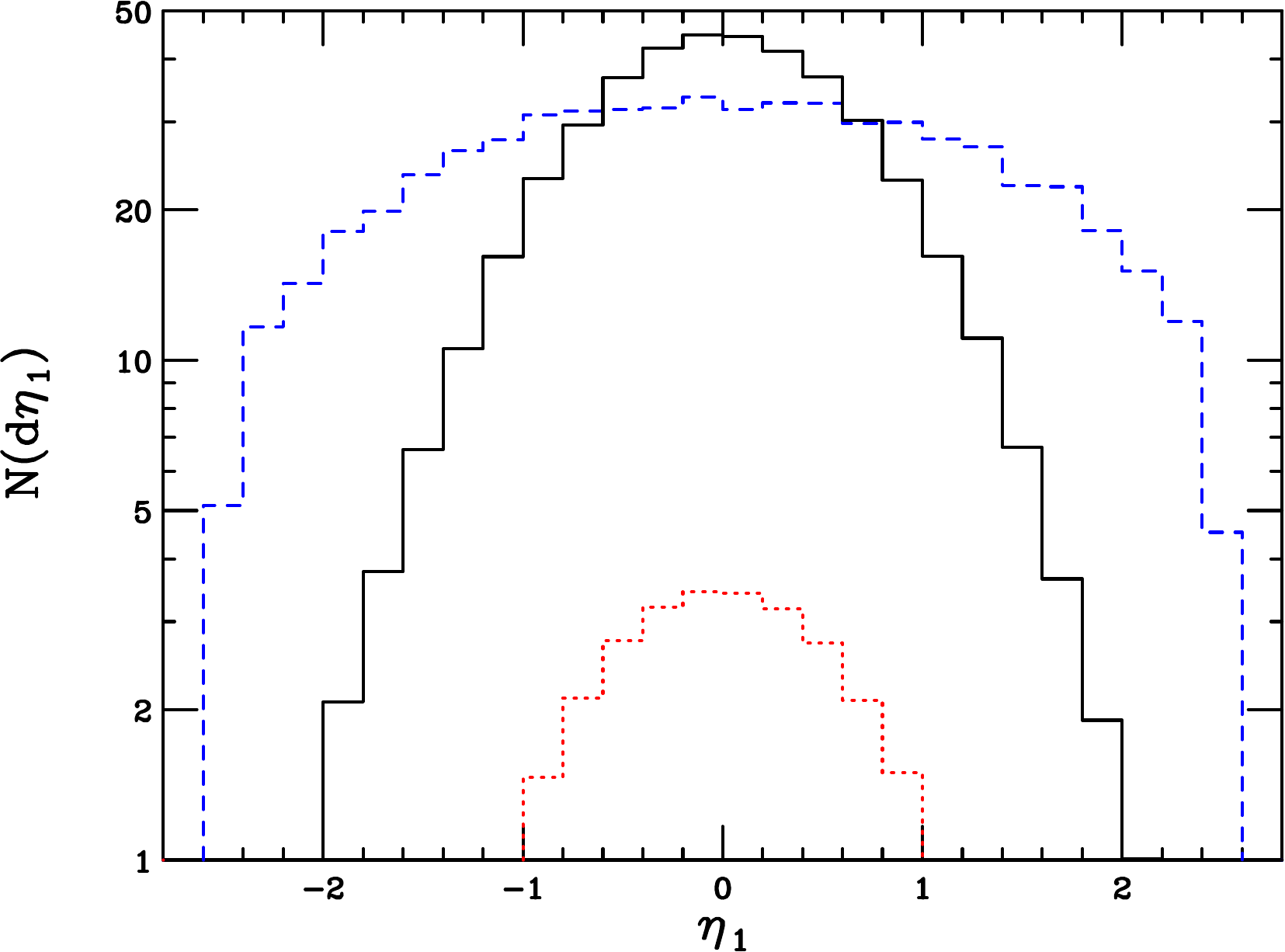}}\hspace{.1cm}
\subfigure[]{\includegraphics[width=0.450\textwidth]{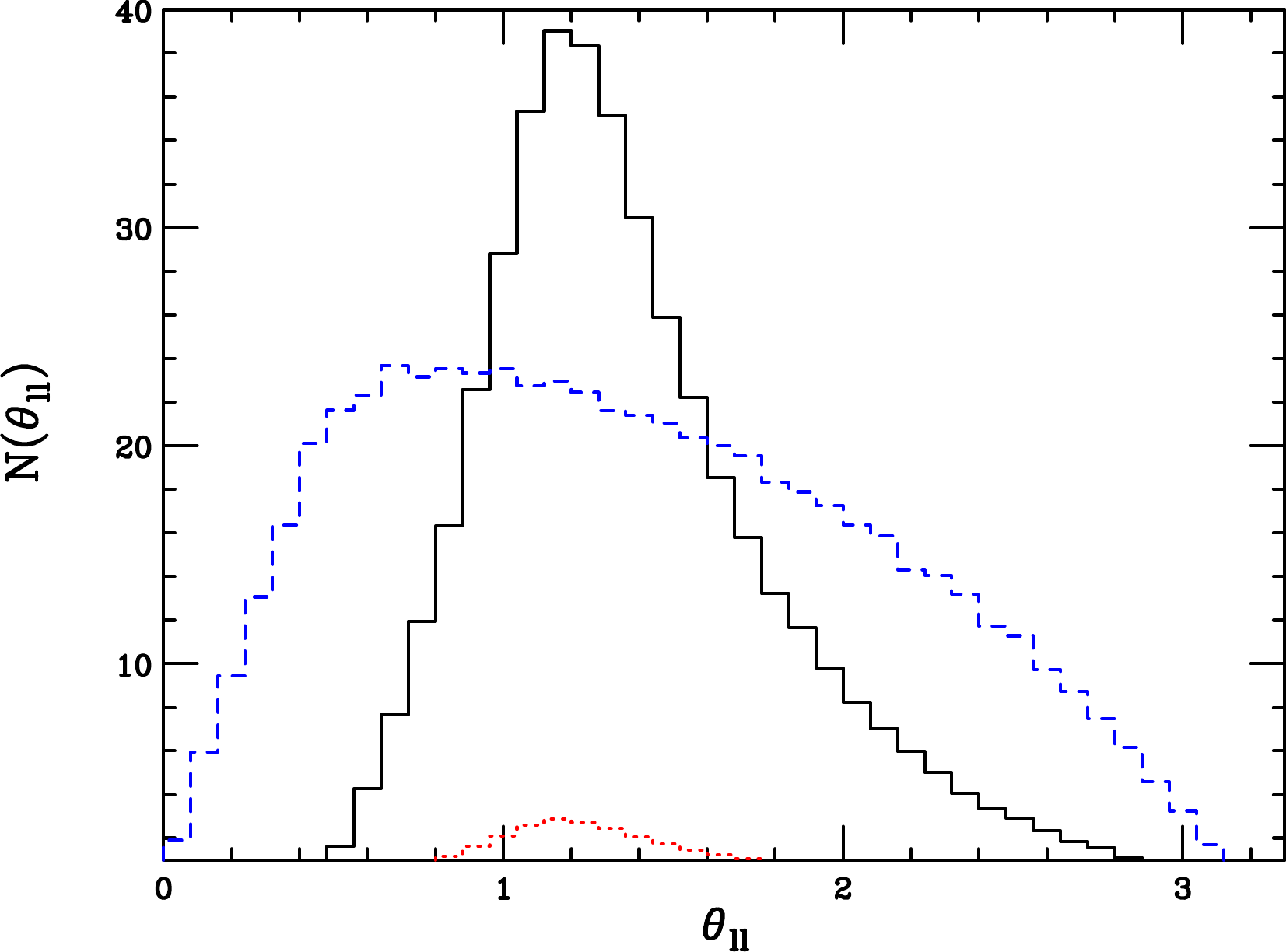}}}
\caption{Distributions of the transverse momentum of the hardest
  (a) and next-to-hardest lepton (b), same-sign lepton invariant mass (c),
  invariant opening angle between the two hardest leptons (d), rapidity
  of the leading lepton (e), polar angle between same-sign leptons (f).
  The solid blue histograms are the spectra yielded by vector bileptons,
  the red dots correspond to scalar doubly-charged Higgs bosons,
the blue dashes to the $ZZ$ Standard Model background.}
 \label{results}
\end{figure}

Besides total cross sections and significances, computed employing the
foreseen number of events, it is instructive studying some final-state
observables, in order to understand how one can possibly
detect (mostly vector-like) bileptons at LHC.
In Fig.~\ref{results} we present the transverse momentum of the
hardest and next-to-hardest lepton ($p_{T,1}$ and $p_{T,2}$),
the invariant opening angle between them ($\Delta R$),
the rapidity of the hardest lepton ($\eta_1$),
the invariant mass ($m_{ll}$) and the
polar angle ($\theta_{ll}$) between same-sign leptons.
In any figure, the results corresponding to
$YY$ (black solid histogram) $HH$ (red dotted histogram)
and $ZZ$ production (blue dashed histogram) are displayed.
Unlike Ref.~\cite{cccf}, where all our spectra were
normalized to 1, in Fig.~\ref{results}
all distributions are normalized in such a way that the height
of each bin, such as $N(p_T)$, yields the expected number
of events for such values of $p_T$, $\eta$, $\Delta R$,
$\theta$ and $m_{ll}$ for a luminosity of
300~fb$^{-1}$ and $\sqrt{s}=13$~TeV.

As one could foresee from the very cross section
and significance evaluations,
the general feature of such spectra is that the 331 signal
can be discriminated from the $ZZ$ background, while
it is not possible to detect doubly-charged Higgs pairs as the
leptonic spectra are always significantly below those yielded by $ZZ$
background and $YY$-pair production. 
As for the transverse momenta ($p_{T,1}$ and $p_{T,2}$),
the $ZZ$ distributions are rather sharp and peak at low $p_T$, while
those yielded by the $HH$ and $YY$ bileptons are much broader and
peak about 1 TeV ($p_{T,1}$) and at roughly 700 ($p_{T,2}$, $YY$)
and 800 GeV ($p_{T,2}$, $HH$).
Such a result should have been expected, since the
$Z$ decays into different-sign, while $Y^{\pm\pm}$ and $H^{\pm\pm}$
into same-sign electrons and muons.
As anticipated, for every value of
$p_T$, the $HH$ spectrum is well below the $YY$ one.

Regarding the rapidity ($\eta_{1,l}$) distribution of the leading
lepton, the 331-model spectra are narrower than the background
and yield a larger event fraction around $\eta_{1,l}\simeq 0$.
Once again, since the $ZZ$ background and the $YY$ signal predict
a number of events of a similar order of magnitude, while
those due to scalar pairs are much lower, the rapidity spectrum
can be useful to detect possible vector bileptons, but not
to separate them from doubly-charged Higgses.

As for the polar angle between same-sign leptons $\theta_{ll}$,
the $YY$-inherited spectrum is peaked around $\theta_{ll}\simeq
1.2\simeq 70^\circ$, while the background is much broader
and maximum at about $\theta\simeq 0.7\simeq 40^\circ$;
the Higgs-like signal is instead negligible.

Concerning the same-sign lepton invariant mass $m_{ll}$,
it is of course easy to discriminate the 331 signals, peaking at
$m_{ll}\simeq 900$~GeV, from the $Z$-pair background, which is
instead a broad distribution, significant up to about 350 GeV
and maximum around 70 GeV. As to the bileptons, both 
invariant-mass spectra are pretty narrow, which 
reflect the fact that $Y^{++}$ and $H^{++}$ have widths
roughly equal to 7 GeV and 400 MeV, respectively.

The distribution of the invariant opening angle $\Delta R$ between
the hardest and next-to-hardest leptons is rather broad for the
background, significant for $0<\Delta R<6$, while
$YY$ pairs yield a distribution in the range  $1<\Delta R<5$,
which, for $\Delta R\simeq 3$, even leads to more events that
the background. The $HH$ signal is possibly visible only for
$2<\Delta R<4$, but even in this range it is negligible
with respect to the SM background and the $YY$ signal.

Our conclusion is therefore that, as already argued in \cite{cccf},
the LHC will be sensitive to the spin-1 bileptons of the 331 model
already at 13 TeV and 300~fb$^{-1}$, and even more in the high-luminosity
regime. If the LHC does not see any bilepton, it may mean either that
bileptons do not exist or that they are scalars, since we have
shown that the production of doubly-charged Higgs bosons in the 331 model is
overwhelmed by the SM background, as well as by vector bileptons.

\section{Discussion}

We explored scalar ($H^{\pm\pm}$) and vector ($Y^{\pm\pm}$)
bileptons in the framework of the 331 model,
which has the appealing features of predicting anomaly
cancellation and treating differently the third quark family with respect
to the first two.
We focused on the family embedding in the minimal 331 model
and paid special attention to its scalar content, and especially
to the sextet sector, whose presence enriches the particle
spectrum and, in particular, leads to the prediction of
doubly-charged scalar Higgs bosons.
Such scalar bileptons can possibly compete
with vector bileptons as a source of same-sign lepton
pairs at the LHC.
In fact, 
previous investigations on vector bileptons, such as \cite{nepo},
had put exclusion limits on the mass of $Y^{\pm\pm}$ exploiting
the experimental searches for scalar $H^{\pm\pm}$, as if the
bilepton spin had a negligible effect on the expected and observed limits
at 95\% confidence level.

We implemented the 331 model, including the new sextet content,
in a full Monte Carlo simulation framework and chose
a benchmark point of the parameter space, consistently
with the present exclusion limits on BSM physics.
We studied jetless events, with
doubly-charged vectors and scalars produced at the LHC
in Drell--Yan  interactions mediated by photons, $Z$, $Z'$
and neutral Higgs bosons,
as well as in processes where exotic quarks
are exchanged in the $t$-channel.

It was found that vector bileptons can be produced
with a
significant cross section already in the present LHC run at 13 TeV and that they
can be easily discriminated from the SM background, by exploring distributions
like the lepton transverse momentum, invariant mass, rapidity or invariant
opening angle. The production of doubly-charged scalars is in principle
interesting, but, because of the helicity suppression,
its cross section is too low for them
to be substantially visible at the LHC and separable from
the background and the vector-bilepton signal.

Our study therefore confirms that, as already anticipated in a previous
analysis, the production of vector-bilepton pairs is the striking
feature of the 331 model
and we believe that, given the large cross section and easy 
separation from the background, with a significance
between $6\sigma$ and $9\sigma$,
it should deserve
a full experimental search. As for doubly-charged scalars, although
the 331 framework discussed in this work leads to a too low production
cross section at the LHC, in order to achieve angular-momentum
conservation, 
we plan to explore how much this 
conclusion depends on the actual setup for the family embedding
and on the choice of the reference point, and whether there could
be other realizations of the model yielding a visible
LHC rate even for scalar bileptons.
Besides, it will be very interesting to study the phenomenology of
the exotic quarks predicted by our 331 model and the LHC
significance reach, especially in the high-luminosity and
high-energy phases.
This is in progress as well.

\section*{Acknowledgement}

We acknowledge Antonio Sidoti for discussions on the cuts implemented
in Eq.~(\ref{cuts}) and on the overlap-removal algorithm
implemented by the ATLAS Collaboration.
This work is partially supported by INFN `Iniziative Specifiche' QFT-HEP
and ENP.

\appendix

\end{document}